\documentclass[aps, pra, preprint, groupedaddress, amsfonts,
               amsmath, amssymb, showpacs, nofootinbib]{revtex4-1}
\usepackage{microtype}
\usepackage{graphicx}
\usepackage{epstopdf}
\usepackage[utf8]{inputenc}
\usepackage[T1]{fontenc}
\usepackage[usenames,dvipsnames]{xcolor}
\usepackage{hyperref}

\usepackage{color}

\begin{document}

\title{Properties of the Sachs electric form factor of the proton on the basis of recent $e-p$ scattering experiments and hydrogen spectroscopy}
\author{Marko Horbatsch} 
\email[]{marko@yorku.ca}
\affiliation{Department of Physics and Astronomy, York University,Toronto, Ontario, Canada M3J 1P3}
\date{\today}							

\begin{abstract} 
Recently published data on the Sachs electric form factor by the PRad collaboration (Nature {\bf 575}, 147-151) are analyzed to investigate their consistency with the known proton charge radius from muonic and electronic hydrogen spectroscopy, as well as theoretical predictions from dispersively improved chiral perturbation theory.
It is shown that the latter is fully consistent with the data, and pointers are given how future $e-p$ scattering experiments can lead to an improvement of our knowledge of the form factor in the low-momentum-transfer regime.
\end{abstract}

\maketitle
\section{Introduction}
The so-called proton radius puzzle appears to be resolved. The puzzle emerged in 2010 when a muonic hydrogen measurement of the $n=2$ Lamb shift~\cite{Antognini417} found that the very accurately
measured proton charge radius $R_E=0.8409(4) \ \rm fm$ (or using a more conservative, model-independent analysis $R_E=0.8413(15) \ \rm fm$)
 disagreed with previous measurements of regular atomic hydrogen intervals~\cite{RevModPhys.88.035009}, quoted in 2014 as $R_E=0.8751(61) \ \rm fm$, as well as the state-of-the-art Mainz $e-p$ scattering experiment (MAMI)~\cite{PhysRevLett.105.242001,PhysRevC.90.015206} with a result of $R_E=0.879(8) \ \rm fm$. The radius $R_E$ enters the spectroscopic analysis via the slope of the Sachs electric form factor at zero momentum transfer squared $Q^2$. The fact that hydrogen spectroscopy and $e-p$ scattering
are determining the same quantity is documented well in the literature~\cite{PhysRevC.99.035202}. 

Since then, numerous efforts were undertaken to resolve the puzzle: {\it (i)} measurements on muonic deuterium~\cite{Pohl669} combined with the isotope shift, {\it (ii)} a fluorescence-based determination of the regular hydrogen $2S-4P$ fine structure intervals~\cite{Beyer79}, and {\it (iii)} a high-accuracy measurement of the Lamb shift in regular hydrogen~\cite{Bezginov1007}  
all pointed to a confirmation of the muonic hydrogen result; on the other hand  {\it (iv)} a high-precision re-measurement of the $1S-3S$ interval by the Paris group~\cite{PhysRevLett.120.183001} continued to support the original higher value for the charge radius; this latter work is being contested by current fluorescence-detection work in Garching, which is
achieving substantially higher precision.

In more recent $e-p$ scattering experiments both the Mainz group through a different method, based on intermediate-state radiation (ISR)~\cite{MIHOVILOVIC2017194} found consistency with the muonic charge radius (albeit with insufficient accuracy to make a strong case, so far), 
as did the PRad collaboration~\cite{PradNature} which employed a gas jet target and measured projectile deflections directly. The situation still has the attention of both the
spectroscopy and scattering communities, but the originally spread ideas that there could be new physics, i.e., that muons and electrons might behave differently
have been damped by these developments.

The significance of resolving the puzzle is not just academic, i.e., eventually, lattice gauge calculations within quantum chromodynamics will be able to compute at least certain aspects of the electric and magnetic form factors, and it will be good to have a solid understanding of the charge and current distributions of the proton based on experimental data. 
In addition, the determination of the charge radius leads to a significant change in the Rydberg constant which links atomic units to SI~\cite{PhysRevA.93.022513,Beyer79}, and settling on it and on the proton charge radius
 opens the possibility for further tests of quantum electrodynamics in atomic hydrogen.
In its most recent update CODATA has adopted the small (muonic) radius value of $0.8414(19)$ fm~\cite{NIST_constants}. When more spectroscopic information supporting the small radius value comes in ($1S-3S$ measurement from Garching, other intervals, as well as deuterium measurements), the uncertainty may decrease in the future.

The analysis of the publicly available and very extensive  MAMI $e-p$ scattering data was challenged by a number of researchers, and much of the controversy focused on the question to what extent one could determine the moments of the proton charge distribution reliably by fitting polynomials (or other functions) to the form factors as a function of $Q^2$, vs a conformal mapping approach that takes care of the branch cut that arises in the analytic continuation of the form factor at the two-pion threshold, or $Q_0^2\approx-0.078 \rm \ GeV^{2}$~\cite{PhysRevD.82.113005} (we make use of $c=1$ units throughout this work). 
In this approach the $Q^2$-dependence of the data is mapped onto a dimensionless variable, usually called $z$, such that they appear at $0<z<1$, while the branch cut is mapped to $z=-1$. 

The purpose of this work is to demonstrate to what extent PRad data for the electric Sachs form factor $G_E(Q^2)$, or their mapped counterpart $G_E(z)$ are consistent with the small radius value $R_E=\sqrt{\langle r^2 \rangle}$ combined with information about the next moment, i.e., $\langle r^4 \rangle$ in accord with the expansion
\begin{equation}
G_E(Q^2)= 1 - \frac{1}{3!} \langle r^2 \rangle Q^2 + \frac{1}{5!} \langle r^4 \rangle  Q^4 - \frac{1}{7!} \langle r^6 \rangle  Q^6 + ...
\end{equation}
Our intent is not to fit the PRad data, but to use their extracted electric form factor data. The PRad analysis is based on robust fitting techniques~\cite{PhysRevC.98.025204}.
In this work we present an alternative analysis technique by following up on a recent suggestion~\cite{HAGELSTEIN2019134825}.

Different analyses of the MAMI data led authors to believe that the fourth moment  $\langle r^4 \rangle$ should be of order $2.0 \rm \ fm^4$ or bigger~\cite{DISTLER2011343,atoms6010002}. Pure chiral perturbation theory (with pions, or
with pions and Delta resonances as degrees of freedom) predicts values below $1.0 \rm \ fm^4$~\cite{PhysRevC.95.035203}. The recently developed approach of dispersively improved higher-order chiral perturbation theory~\cite{PhysRevC.97.055203} in a first version made  a prediction of $1.43(27) \rm \ fm^4$, and on higher moments as well, but no prediction for the second moment, i.e., the charge radius. The electric and magnetic charge radii of the nucleons are taken as input (with uncertainties) from the Particle Data Group. 
 
In the most recent version of the improved chiral perturbation theory, where the magnetic form factors are taken into account to the same order as the electric ones,
a tighter uncertainty range is obtained~\cite{ALARCON2018373}, i.e., the current prediction for the fourth electric moment of the proton is $1.537(65) \rm \ fm^4$,
and it is shown that the electric and magnetic form factors agree well with scattering data (which are dominated by the MAMI data set). 
One might have a small concern that the input for the magnetic charge radius of the proton is biased
against the MAMI result, i.e., the uncertainty range is assumed to be $0.83 < R_p^{\rm mag} < 0.875 \ \rm fm$, which excludes the MAMI result~\cite{PhysRevC.90.015206} 
of about $R_p^{\rm mag}=0.80(2) \ \rm fm$ (with appropriate two-photon exchange corrections taken into account in the data analysis).

Thus, the question arises whether the PRad data (and future $e-p$ scattering data) have strength on resolving the discrepancy concerning $\langle r^4 \rangle$
between earlier MAMI analyses and the theoretical prediction from Ref.~\cite{ALARCON2018373}, and the present work deals with this issue. We present analyses both
in terms of $Q^2$ and the conformal mapping variable $z$.

The conformal mapping mentioned above for the choice of expansion point $z_0=0$ corresponding to $Q^2=0$ is defined by
\begin{equation}
z=\frac{\sqrt{Q^2+t_c}-\sqrt{t_c}}{\sqrt{Q^2+t_c}+\sqrt{t_c}}
\end{equation}
where $t_c= 4 m_\pi^2$ is defined in terms of the pion mass. The inverse map
\begin{equation}
Q^2=\frac{4 t_c z}{(1-z)^2}
\end{equation}
demonstrates how the range $0 < Q^2 < \infty$ is mapped onto $0<z<1$ with a linear relationship for small $Q^2$.

The two-pion threshold in photon-nucleon scattering is only the first such threshold, i.e., there is also a three-pion threshold at $t_{3\pi}= 9 m_\pi^2$. This threshold is mapped onto some place on the unit circle $|z|=1$, so it can be argued that it is also out of harm's way. It can be demonstrated that analytically computed form factors in chiral perturbation theory do have Taylor series in the $Q^2$ variable with the radius of convergence given by $t_c$, i.e., they are of very limited range, and that this problem can be cured by considering Taylor expansions in $z$. Thus, we also consider the following expansion
\begin{equation}
G_E(z) = 1 - p_1 z +p_2 z^2 + ...
\end{equation}
The moments of the form factor again are related to the expansion coefficients, e.g.,
\begin{equation}
R_E=\sqrt{\langle r^2 \rangle} = \sqrt{\frac{3 p_1}{2t_c}} \ .
\end{equation}

Attempts to fit the MAMI data to high-order polynomials in $z$ have a tendency to result in larger values of the proton charge radius~\cite{PhysRevD.92.013013,PhysRevD.91.014023}. As a result we consider alternatives
to the polynomial expansions (1) or (4) in the form of Pad\'e rational functions that agree with low-order polynomials of form (1) or form (4) respectively.
It is straightforward (e.g., using Mathematica's function PadeApproximant) to define functions such as Rational[1,1] (used in Ref.~\cite{PradNature}) or higher-order versions which can be constrained by using information about the moments from dispersively improved chiral perturbation theory.

Recently an interesting proposal was made by Hagelstein and Pascalutsa~\cite{HAGELSTEIN2019134825} to analyze the from factor data by taking the logarithm. One can turn the expansions (1) or (4) into expressions that yield $Q^2$-dependent (or $z$-dependent) radius functions, e.g.,
\begin{equation}
R_E(Q^2) =\sqrt{ -\frac{6}{Q^2}\ln{G_E(Q^2)} } \ .
\end{equation}
Arguments are provided for the property of the true radius $R_E \equiv R_E(0) \le R_E(Q^2)$, although it is not clear whether a bounding property is all that meaningful
when dealing with data that have statistical and systematic errors. 
The arguments are based on relating $G_E(Q^2)$ to the three-dimensional charge density, which is conceptually being questioned~\cite{PhysRevC.99.035202}. 
Note that taking the logarithm amplifies errors at small $Q^2$. Nevertheless, we find this tool useful
to discover inconsistencies in the data, particularly at low $Q^2$. It was argued that this analysis is less dependent on the normalization constants,
which are considerable factors of uncertainty in the extraction of radius values, from both the MAMI and PRad data sets.

For the conformal mapping version no bounding property has been derived. The definition for the function $R_E[z]$ follows by analogy, i.e.,
\begin{equation}
R_E(z) = \sqrt{-\frac{3}{2 t_c z}\ln{G_E(z)}} \ .
\end{equation}
From the results shown further below one may be led to the conjecture that $R_E(z)$ approaches the true radius value $R_E(0)$ from above, but also that the $z$ expansion
introduces a strong dependence on the $Q^2$ range of data included in the analysis.

\section{Data Analysis for $G_E(Q^2)$ and $G_E(z)$}

\begin{figure}
\begin{center}
\resizebox{0.6\textwidth}{!}{\includegraphics{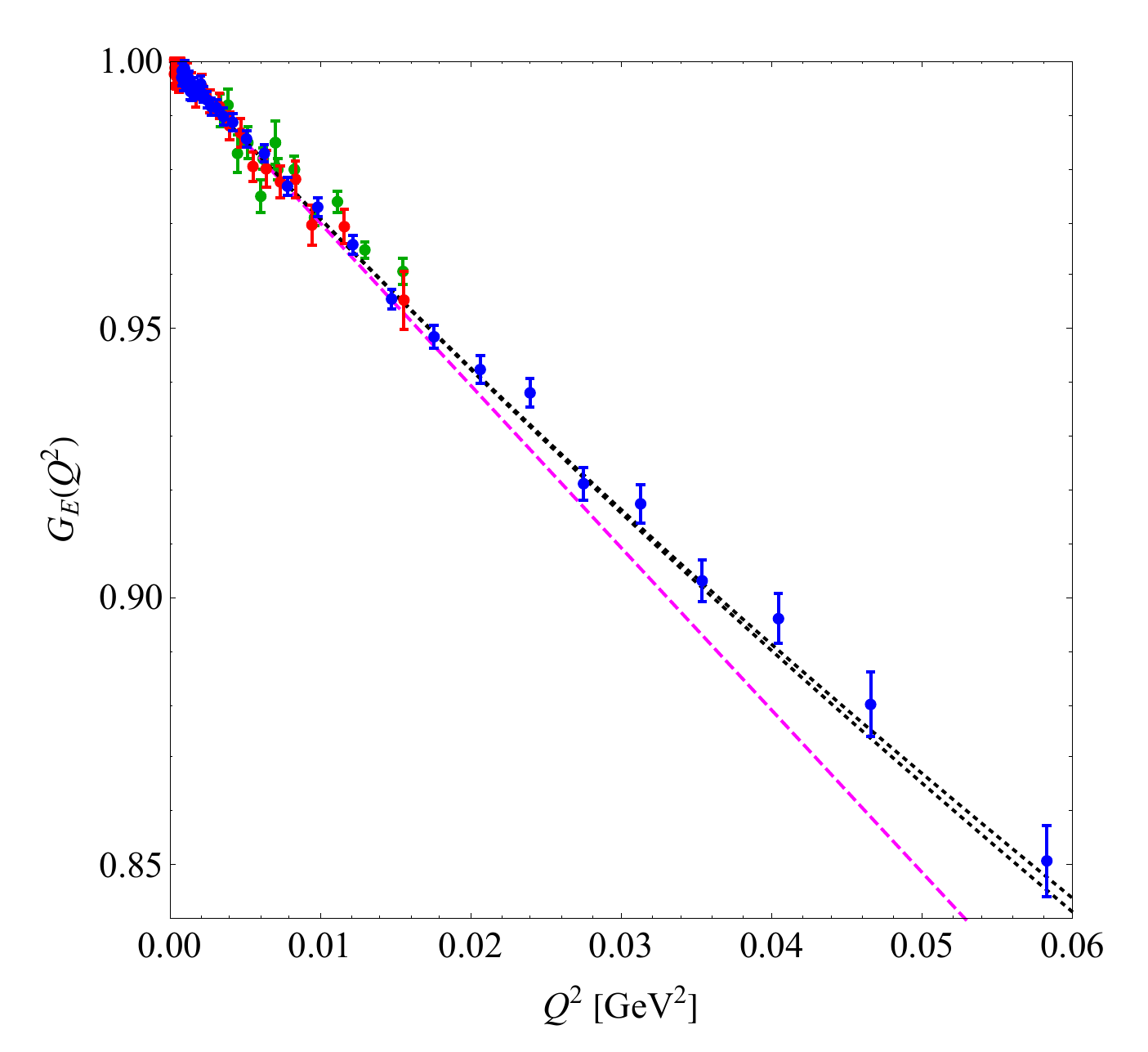}}  

   \caption{The Sachs electric form factor for the proton as a function of momentum transfer squared: $G_E(Q^2)$. Red and blue data points are from PRad~\cite{PRadWiki}
   for $E=1.1 \ \rm and \ 2.2 \ GeV $ respectively
   with statistical and systematic error estimates added in quadrature, while the green data points
   are from the Mainz ISR experiment. The magenta dashed line shows the truncated function at first order with the muonic hydrogen value used in (1). The black dotted curve pair
   shows the result of the rational function~(8) determined such that its slope at $Q^2=0$ corresponds to the muonic hydrogen radius $R=0.841 \ \rm fm$, while the curvature at $Q^2=0$
   corresponds to the error band established as $1.47 \le \langle r^4 \rangle \le 1.60 \ \rm fm^4$ as predicted by dispersively improved chiral perturbation theory~\cite{ALARCON2018373}.}
   \label{fig:fig1}
\end{center}
\end{figure}

In Fig.~1 we show the  PRad~\cite{PradNature} data with statistical and systematic errors added in quadrature~\cite{PRadWiki}, and also some
low-$Q^2$ ISR data from Mainz~\cite{MIHOVILOVIC2017194} for comparison.
It turns out that the PRad results from the 1.1 GeV beam energy which correspond to very low values of $Q^2$ are not very useful in constraining 
the proton radius, and a similar comment can be made about the preliminary ISR data (a future data run of the ISR experiment in Mainz is planned~\cite{ISRfuture}).
The 1.1 GeV data from PRad are, however, important in determining the normalization of the data, and only a small deviation from unity in the normalization constants for the two energy
runs is required to achieve a satisfactory fit to both data sets~\cite{PradNature}.
On the basis of Fourier transforms of the proton charge density it was argued in Refs.~\cite{PhysRevC.95.012501,atoms6010002} that the sensitivity range in the data to the proton charge radius (or $\langle r^2 \rangle$) is
about $0.01 < Q^2 < 0.04 \ \rm \ GeV^2$, while sensitivity to the fourth moment $\langle r^4 \rangle$ would fall to the right of that interval with maximum
sensitivity at around $Q^2=0.08 \ \rm GeV^2$ when the next higher moment would begin to play a significant role.

The dashed straight line shows the consistency of the low-$Q^2$ data with the CODATA2018 radius value~\cite{NIST_constants}. It also shows that the $2.2 \ \rm GeV$ PRad data contain information
about the fourth moment. 
The dotted curves are obtained as follows: a three-parameter Pad\'e approximant Eq.~(8) was obtained from a polynomial in $Q^2$, such as Eq.~(1), where
the coefficient linear in $Q^2$ was fixed to correspond to the muonic hydrogen radius value $R_p=0.841 \ \rm fm$, the coefficient with $Q^4$ was chosen to correspond
to either the higher or lower bracket value of the dispersively improved chiral perturbation theory prediction~\cite{ALARCON2018373}, while the coefficient with $Q^6$
was chosen as the center value of this theoretical prediction, i.e., corresponding to $\langle r^6 \rangle = 8.7 \ \rm fm^6$. Incorporating the theoretical uncertainty in this latter term does not lead to noticeable changes in the curves for the range of $Q^2$ shown in Fig.~1.
These are not fits, they are the results of combining spectroscopy (radius value) with state-of-the-art theory.
Including the $\langle r^6 \rangle$ moment, and therefore using a three-parameter rational function ensures that the form factor function is valid somewhat beyond the data
range of the PRad experiment.

The function pair shown in Fig.~1 is defined as 
\begin{equation}
G_E(Q^2) = \frac{1-a_1 Q^2}{1-b_1 Q^2-b_2 Q^4}
\end{equation}
and is applicable for $Q^2 < 0.078 \ \rm GeV^2$ to avoid issues caused by the two-pion threshold (poles may occur for larger $Q^2$ in Eq.~(8)). 
All three parameters are pre-determined as described above.
In the range relevant for the PRad data (and up to about $\rm 0.1 \ GeV^2$) the trace of the functions agrees very well with the function given in Fig.~3 of Ref.~\cite{ALARCON2018373}.
The coefficients $\{a_1, b_1, b_2 \}$ are easily obtained in a symbolic computation package which determines a Pad\'e approximant from a given polynomial.

The agreement with the PRad data is very good. 
The quality of the predicted higher moments from Ref.~\cite{ALARCON2018373} was previously tested on larger $e-p$ scattering data sets~\cite{PhysRevC.99.044303}.
Here we show that the PRad data set lends strong support to these theoretical results. A detailed investigation of the MAMI data 
along similar lines is also in progress, particularly in order to assess
its determination of the magnetic charge radius whose small value is in conflict with other experiments and their analysis~\cite{PhysRevD.98.030001,PhysRevD.90.074027}.
Extending the tools presented here will also determine the fourth electric moment to higher precision than the prediction of Ref.~\cite{ALARCON2018373}.

\begin{figure}
\begin{center}
\resizebox{0.6\textwidth}{!}{\includegraphics{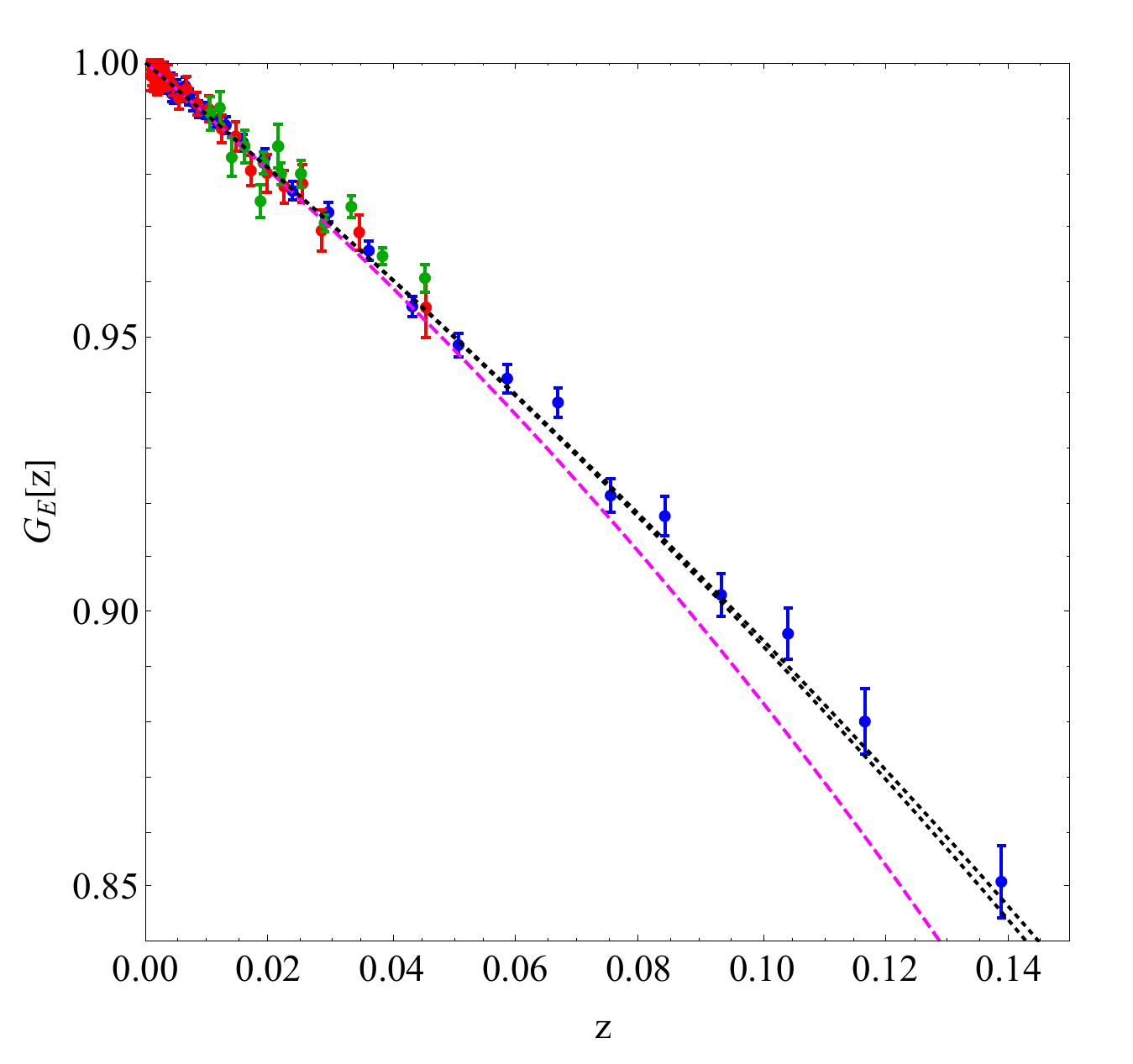}}  
   \caption{The Sachs electric form factor for the proton as a function of the conformal mapping variable, i.e., $G_E(z)$. The experimental data are labeled as in Fig.~\ref{fig:fig1}. The magenta dashed line shows the truncated function at first order with the muonic hydrogen value used in (1) and converted to become a function of $z$, i.e., it is non-linear in $z$. The black dotted line represents the error band as established by dispersively improved chiral perturbation theory, cf. Fig.~1, but follows
   from a rational function in the $z$ variable analogous to Eq.~(8) and obeys the derivative conditions at $Q^2=0$, which corresponds to $z=0$.}
   \label{fig:fig2}
\end{center}
\end{figure}

In Fig.~2 the data are presented as a function of the conformal mapping variable $z$. Even though it looks like a complication, in that the simplest form factor function truncated at
order $Q^2$ becomes a curve, the idea of using this representation is rooted in the fact that there is no difficulty with the radius of convergence for a power series in $z$.
In fact, the conformal mapping allows one to construct the entire form factor function from a given set of moments, something that is not possible in the $Q^2$ variable
due to the limited radius of convergence for the series in $Q^2$. 
For the range
of the PRad data this is not important, because they do not reach beyond the critical point, i.e., $0.078 \ \rm GeV^2$. 

The conclusions to be drawn from a three-parameter
Pad\'e function in $Q^2$ (Fig.~1), or in $z$ (Fig.~2) are basically the same: the experimental data cannot be used directly at lowest $Q^2$ or $z$ to measure the
derivatives of the form factor. However, incorporating such derivatives on the basis of hydrogen spectroscopy (for $R_E$) and dispersively improved higher-order 
chiral perturbation theory (for the higher moments) demonstrates
consistency with the PRad results. It is very likely that such a procedure will be required also for future low-$Q^2$ experiments for $e-p$ or $\mu-p$ scattering.

\begin{figure}
\begin{center}
\resizebox{0.6\textwidth}{!}{\includegraphics{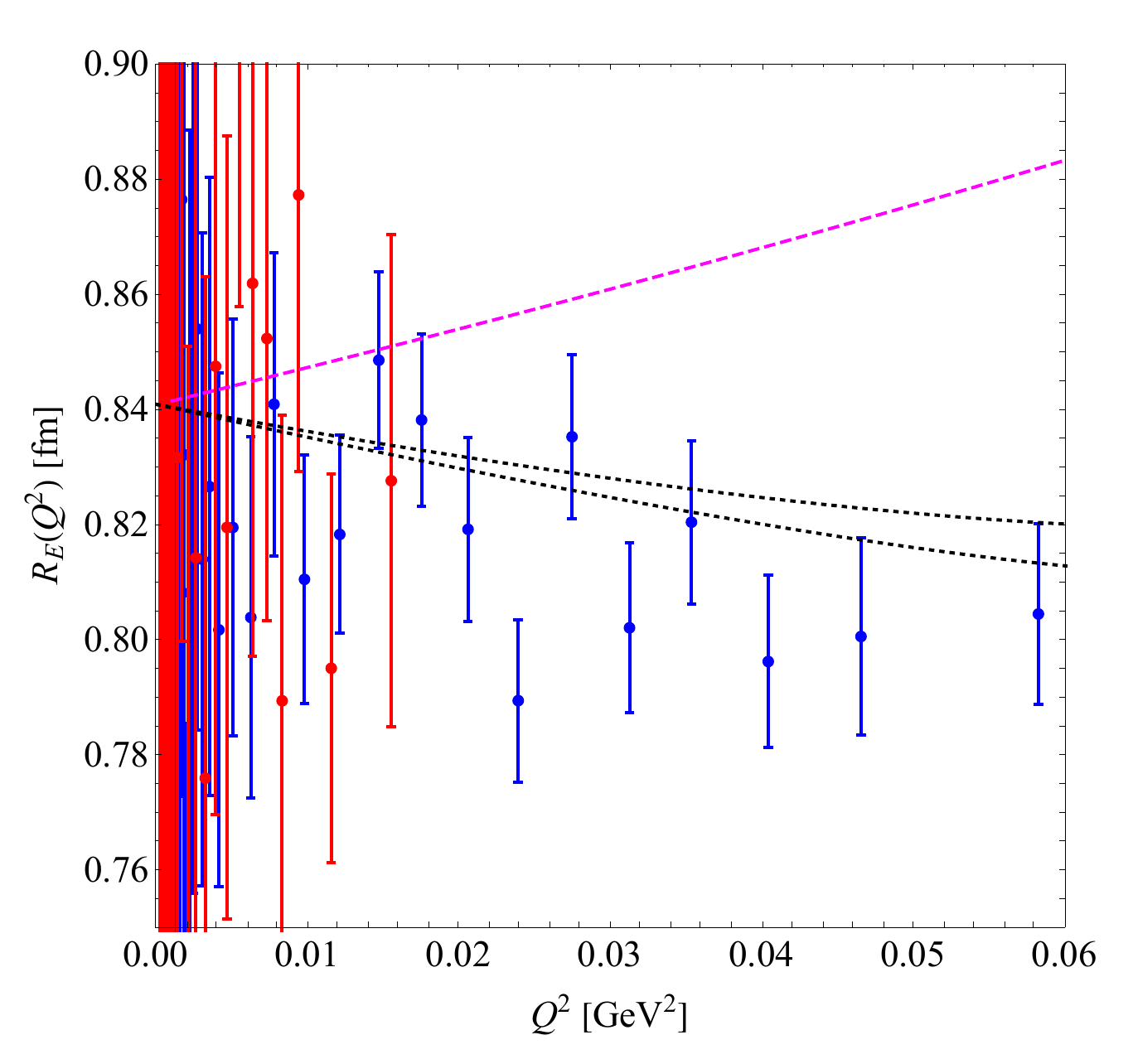}}  

   \caption{The proton electric charge radius function $R_E(Q^2)$ in fm obtained from eq.~(6). The data points from the PRad experiment are shown in red and blue for the
   1.1 and 2.2 GeV data runs respectively, while the dotted curves correspond to the equivalent result in Fig.~1 which is a prediction based on the spectroscopic value of the charge radius and a theoretical prediction for the higher moments.
   The dashed magenta curve corresponds to the straight-line result in Fig.~\ref{fig:fig1}.}
   \label{fig:fig3}
\end{center}
\end{figure}

\section{Data Analysis for $R_E(Q^2)$ and $R_E(z)$}

In Fig.~3 we show the result of the transformation given in eq.~(6). The mapping to the function $R_E(Q^2)$ scales up the errors for low $Q^2$, and it becomes evident
that the $2.2 \ \rm GeV$ data set covers the range required in order to pin down the radius value without being affected too much by the contribution from $\langle r^4 \rangle$.
Interestingly, the data points below $0.01 \ \rm GeV^2$ do not contribute towards a strong statement (as anticipated in Ref.~\cite{atoms6010002}). This is not immediately
obvious from Fig.~1. The dashed curves correspond to our theoretical result in Fig.~1, i.e., the form of $G_E(Q^2)$ is a known analytic function and can be treated
in the sense of Ref.~\cite{HAGELSTEIN2019134825} as being a lower bound to the proton charge radius. The simple linear form factor result (magenta line in Fig.~1)
does not obey the bound, since it does not correspond to a legitimate charge density.
Experimental data with statistical (and systematic) errors, do
not obey bounds, as can be seen from the PRad data, but they agree very well when taking their standard deviation into account.

\begin{figure}
\begin{center}
\resizebox{0.6\textwidth}{!}{\includegraphics{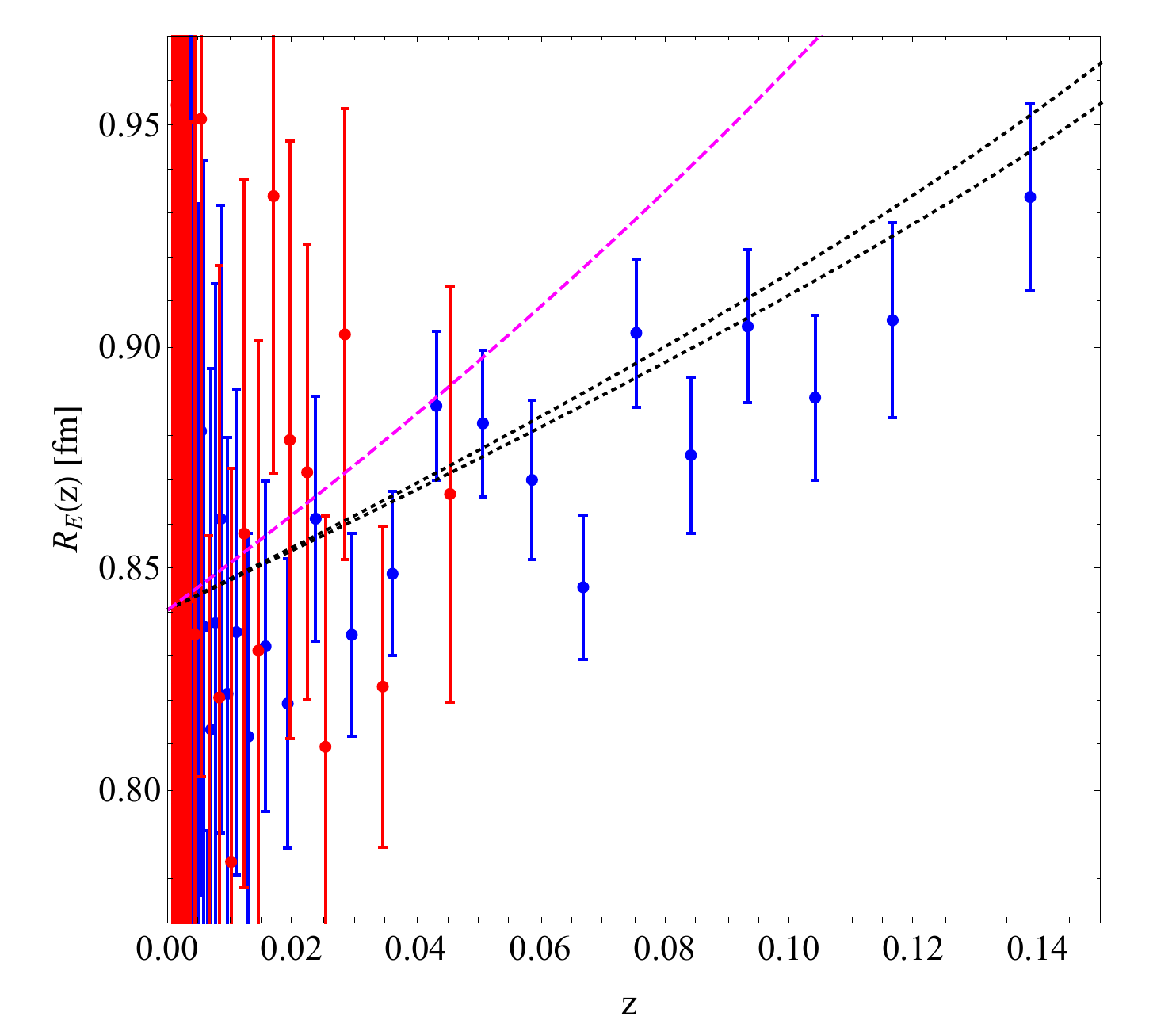}}
   \caption{The proton electric charge radius function $R_E(z)$ in fm obtained from eq.~(7). 
   The PRad data points and the dotted and dashed curves correspond to the equivalent results in Fig.~2.}
   \label{fig:fig4}
\end{center}
\end{figure}

In Fig.~4 we repeat a similar analysis for $R_E(z)$ based on eq.~(7). A rather different picture emerges in this case when looking at
the analytical results based on the Pad\'e form factor as a function of $z$ compared to $R_E(Q^2)$.  
The function $R_E(z)$ rises quickly to large values. It may be an upper bound (we have
no proof, at best a conjecture), but not a very useful one. The data are very consistent with this, but the figure raises the question about the usefulness of fits to $G_E(z)$
with the goal to extract the proton charge radius. The very large values of $R_E$ obtained from $z$-dependent fits to the MAMI data~\cite{PhysRevD.92.013013,PhysRevD.91.014023,PhysRevC.93.015204} may well be connected with the character observed here.
The presentation in the form of $R_E(z)$ again turns out to be useful, since this conclusion would not be drawn from looking at the functions $G_E(z)$ shown in Fig.~2.

\section{Conclusions}

Given that there are a number of lepton-proton form factor measurements at low $Q^2$ in progress, the present work should help with their data analysis. Apart from the
mentioned experiments on muon scattering (MUSE,~\cite{Roy_2020}), and updated MAMI measurements both with magnetic spectrometers but the solid hydrogen target replaced
by a gas het target, as well as to measure
proton recoils using such a target~\cite{Vorobyev_2019} there are
also proposals for measurements in France~\cite{Hoballah_2019} and in Japan~\cite{ELPH}.

Ultimately, one would like to understand not only the low-$Q^2$ dependence of the electric (and magnetic) form factors, but also improve the understanding of how these 
form factors connect to data at high $Q^2$~\cite{YE20188}. Given the difficulty of the MAMI data analysis to connect with the small charge radius one should not only emphasize
the lowest-$Q^2$ region, which is apparently where much of the current efforts will go. 
It will be at least equally important to probe the $Q^2$ regions to the right of $0.08 \ \rm GeV^2$
in order to determine experimentally the higher moments of the electric charge distribution of the proton, and to probe the lowest moments of its 
current distribution to higher precision. This will be of use to the numerical lattice gauge theory community which is working towards improvement on its 
first attempts to determine the form factors from quantum chromodynamics alone~\cite{PhysRevD.100.014509,PhysRevD.99.014510}.

\begin{acknowledgments} Work supported by the Natural Sciences and Engineering Research Council of Canada.
Discussions with Eric Hessels, Douglas Higinbotham, and Weizhi Xiong are gratefully acknowledged.
\end{acknowledgments}

\vfill\eject
 
\bibliography{AnalyzePRadR4}

\end{document}